\documentclass[sigconf,natbib=true]{acmart}
\usepackage{graphicx} 
\usepackage{caption}
\usepackage[linesnumbered,ruled,vlined]{algorithm2e}
\usepackage{subfigure}
\usepackage{multirow}
\usepackage[normalem]{ulem}
\usepackage{balance}
\usepackage[T1]{fontenc}
\usepackage[utf8]{inputenc}
\usepackage{makecell}
\usepackage{enumitem}
\usepackage{amsmath}

\AtBeginDocument{%
  \providecommand\BibTeX{{%
    \normalfont B\kern-0.5em{\scshape i\kern-0.25em b}\kern-0.8em\TeX}}}

\setcopyright{acmcopyright}
\copyrightyear{2018}
\acmYear{2018}
\acmDOI{XXXXXXX.XXXXXXX}

\acmConference[Conference acronym 'XX]{Make sure to enter the correct
  conference title from your rights confirmation emai}{June 03--05,
  2018}{Woodstock, NY}
\acmPrice{15.00}
\acmISBN{978-1-4503-XXXX-X/18/06}

\newenvironment{fequation}{\begin{equation}\footnotesize}{\end{equation}}

\begin{document}

\title{Disentangled Counterfactual Reasoning for Unbiased Sequential Recommendation}

\author{Yi Ren}
\affiliation{%
  \institution{Tencent }
  \city{Beijing}
  \country{China}}  
\email{yiren_bj@outlook.com}

\author{Xu Zhao}
\authornote{Work done while at Tencent.}
\affiliation{%
  \institution{Xiaohongshu }
  \city{Beijing}
  \country{China}}  
\email{zhaoxu2@xiaohongshu.com}

\author{Hongyan Tang}
\affiliation{%
  \institution{Tencent }
  \city{Beijing}
  \country{China}}  
\email{violatang@tencent.com}

\author{Shuai Li}
\affiliation{%
  \institution{Tencent }
  \city{Beijing}
  \country{China}}  
\email{matli@tencent.com}
\renewcommand{\shortauthors}{Yi Ren et al.}

\begin{abstract}
Sequential recommender systems have achieved state-of-the-art recommendation performance by modeling the sequential dynamics of user activities. However, in most recommendation scenarios, the popular items comprise the major part of the previous user actions. Therefore, the learned models are biased towards the popular items irrespective of the user's real interests. Various methods are proposed to address the issue of popularity bias. Traditional methods introduce additional re-ranking steps to augment the coverage of unpopular items or utilize Inverse Propensity Weighting (IPW) to decrease the impact of popular items during model training. Recently, structural causal model-based methods are introduced to achieve superior recommendation performance. Nonetheless, the causal graph proposed before can still be enhanced. Moreover, the design of the existing methods does not leverage the characteristics of sequential model structures, which cannot warrant optimal recommendation performance.

In this paper, we propose a structural causal model-based method to address the popularity bias issue for sequential recommendation model learning. For more generalizable modeling, we disentangle the popularity and interest representations at both the item side and user context side. Based on the disentangled representation, we identify a more effective structural causal graph for general recommendation applications. Then, we design delicate sequential models to apply the aforementioned causal graph to the sequential recommendation scenario for unbiased prediction with counterfactual reasoning. Furthermore, we conduct extensive offline experiments and online A/B tests to verify the proposed \textbf{DCR} (Disentangled Counterfactual Reasoning) method's superior overall performance and understand the effectiveness of the various introduced components. Based on our knowledge, this is the first structural causal model specifically designed for the popularity bias correction of sequential recommendation models, which achieves significant performance gains over the existing methods.  

\end{abstract}



\maketitle
\vspace{-0.2cm}
\section{Introduction}
We have seen widespread adoption of recommender systems, which can support personalized recommendations for various online services, such as social networking, online media, and e-commerce sites. Among the various recommendation algorithms, sequential recommendation models are increasingly prevalent for a variety of scenarios because of the promising recommendation performance and the effectiveness of user modeling. By explicitly reasoning the dynamic correlations between the user's previous actions and successive behaviors, the sequential models can capture and leverage the useful patterns from sequential dynamics to achieve superior recommendation performance. Based on the underlying network architecture, the sequential models can be roughly classified to RNN based\cite{hidasi2015session,hidasi2018recurrent}, CNN based \cite{yuan2019simple,tang2018personalized,tuan20173d} and Self-Attention based \cite{kang2018self,sun2019bert4rec} methods. And all of the aforementioned sequential methods use the user's previous actions as the context so as to accurately predict the next actions. 

\begin{figure}[!htbp]
\setlength{\abovecaptionskip}{-0.0cm} 
\setlength{\belowcaptionskip}{-0.5cm}
	\centering\textbf{}
	\includegraphics[scale=0.45]{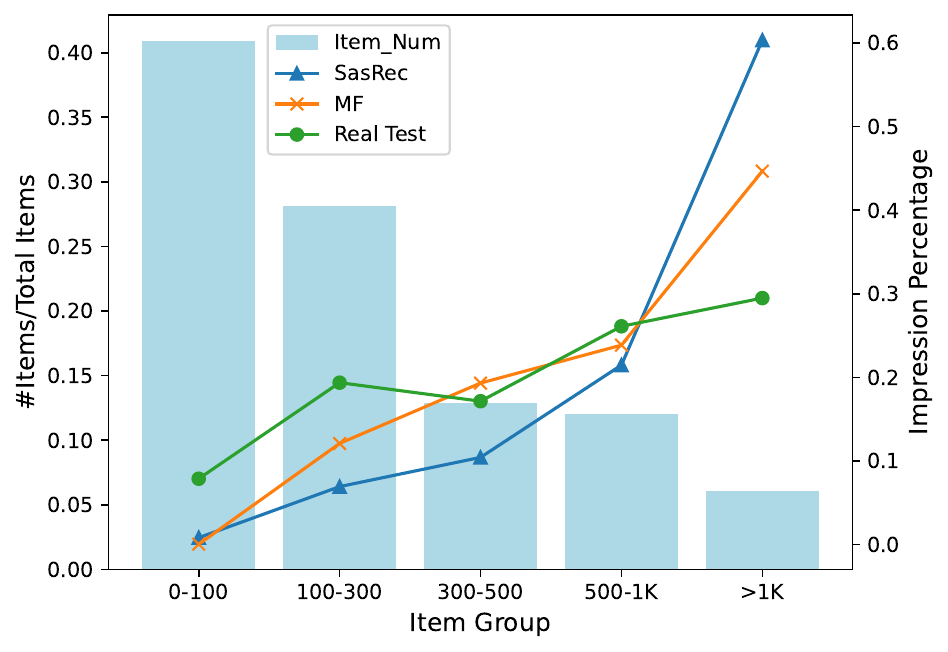}
	\caption{Comparison of Popularity Bias between SASRec\cite{kang2018self} and MF\cite{koren2009matrix} on the ML-1M dataset. Items are grouped by item popularity in the training set. The left vertical axis denotes the item number ratio of each group. And the right vertical axis represents each group's impression exposure percentage during testing. First, SASRec will expose more items in the '>1k' group than MF. Moreover, compared with the real distribution of the test set, both of them over-concentrate on the '>1k' group.}
	\label{fig:rec_dist}
\end{figure}

Most recommendation methods are known to be plagued by the issue of popularity bias \cite{chen2020bias} since the training data is biased towards the popular items. Moreover, the popular items also comprise the major part of the previous user actions, which is the core feature for sequential models. Hence, compared with other recommendation models, the sequential models are more directly affected by such bias and tend to recommend even more popular items without regard to the users' real interests, which can be verified by Figure \ref{fig:rec_dist}.  Based on the previous research \cite{chen2020bias,zhu2021popularity,abdollahpouri2019managing,wang2022unbiased,saito2020unbiased2,saito2020unbiased}, over-concentrating on the popular items will hinder the model's accurate understanding of the user's true preference and likely decrease the recommendation diversity and overall utility of the recommendation service.    

Owing to the prevalence of recommender systems, the problem of popularity bias negatively impacts the experience of billions of users every day. Hence, significant research efforts \cite{zhu2021popularity,abdollahpouri2019managing,wang2022unbiased,saito2020unbiased2,saito2020unbiased,lee2021dual,qin2020attribute,zhu2020unbiased,wei2021model,zheng2021disentangling,chen2022co,zhang2021causal,he2022addressing,wang2021deconfounded} have been put forward in designing effective debiasing algorithms to alleviate the aforementioned issues. First, some methods \cite{zhu2021popularity,abdollahpouri2019managing} introduces additional post-processing and re-ranking steps to increase the exposure of unpopular items and balance the recommendation diversity and utility. Another line of works \cite{schnabel2016recommendations,wang2022unbiased,saito2020unbiased2,saito2020unbiased,lee2021dual,qin2020attribute,zhu2020unbiased} resort to rebalancing the loss weight of different training samples with the inverse propensity weighting (IPW). Although theoretically sound, these methods are not robust to small propensities as small propensities often lead to extremely large sample weights and uncontrollable high variance. Some methods \cite{swaminathan2015self,bottou2013counterfactual,gruson2019offline} are proposed to practically control the variance and alleviate the instability issue at the cost of introducing additional bias. However, the bias-variance trade-off makes the model hard to tune to achieve satisfactory results. 

Recently, structural causal model-based methods \cite{wei2021model,zheng2021disentangling,chen2022co,zhang2021causal,he2022addressing,wang2021deconfounded} are proposed to further advance the debiasing methods for superior performance. The confounding and colliding structures\cite{neuberg2003causality} usually exist in a causal graph reflecting real-world relations of variables. With such unfavorable structures, it is impossible to answer the causal question with correlation-level tools since confounders and controlling colliders bring spurious correlations between treatment and outcome. To cope with this challenge, the structural causal model provides both solid mathematical foundations and friendly calculus for the analysis of causes and counterfactuals. Therefore, it can help to deduce the direct and indirect effects of potential interventions in the counterfactual world. For example, MACR\cite{wei2021model} proposes the framework to counteract popularity bias by carrying out causal reasoning in the counterfactual world. And DICE\cite{zheng2021disentangling} assigns users separate embeddings for interest and conformity to derive generalizable representations by leveraging the colliding structures.  The latest experiment results also prove that these methods can achieve state-of-the-art recommendation performance. However, the utilized causal graph can still be improved. Furthermore, the existing structural causal model-based methods are not specifically designed for the sequential recommendation scenario, which cannot warrant optimal recommendation results in the sequential recommendation setting.    

Accordingly, in this paper, a structural causal model-based method is specifically designed for the sequential recommendation setting to address the popularity bias issue much more effectively. First, we explicitly disentangle the item embeddings to popularity and interest representations and disentangle user preference to conformity and interest representations for generalizable modeling. Second, with the disentangled representations, we identify a fine-grained structural causal graph for general recommenders. Within this causal graph, the popularity and conformity representations exert a direct effect on user interactions. For the indirect effect, the popularity-conformity embeddings and interest embeddings are matched separately and then combined together so as to model to what degree the user interaction is from the match to user interest or conformity. Moreover, considering the superior performance of sequential recommender models, we apply the proposed causal graph to the sequential recommendation and specifically design effective sequential network structures so as to perform counterfactual reasoning for alleviating the popularity bias. Finally, extensive experiments are conducted to verify that this method can achieve superior overall recommendation performance compared with the existing methods. We also perform detailed ablation tests to study the benefit and effectiveness of the various introduced components. 

Compared with MACR\cite{wei2021model}, our model utilizes disentangled representations, which can warrant more generalizable models. In comparison with DICE\cite{zheng2021disentangling}, our disentangling method is much simpler and does not rely on sophisticated data preparation for curriculum learning. And we leverage counterfactual reasoning, which is absent with DICE, for unbiased inference. Furthermore, for accurate bias correction, we utilize separate sequential structures to process the interest and popularity representations of historical interactions to model the user's preference for interest and conformtiy respectively. 
We summarize our main contributions as follows.

\begin{itemize}[leftmargin=.2in]
\item By leveraging disentangled user and item representations, we present a new causal view of the popularity bias in general recommender systems. Disentangled representations contribute to more generalizable and robust modeling\cite{zheng2021disentangling}.
\item We apply this causal graph to the sequential recommendation scenario. For accurate bias correction, we design separate sequential structures based on the interest or popularity representations of historical interactions to precisely model the user's preference for interest or conformtiy.
\item Thanks to the proposed approach, counterfactual reasoning can be performed to compensate for the popularity bias with proper operation in the training and inference stage. 
\item We conduct extensive experiments with multiple real-world datasets over RNN-based, CNN-based, and Self-Attention-based backbone networks to verify and understand the effectiveness of the overall method and each individual component.
\item We perform online A/B testing to prove the method can be practically applied in industry recommender systems to enhance the utility of the recommendation service.

\end{itemize}
\vspace{-0.2cm}
\section{Related Works}

\subsection{Sequential Recommendation}
Traditional recommendation models discard any sequential information and generate recommendations by learning the user's general interests. As a result, their recommendation performance can be limited without considering the rich contextual information of user action histories. Various sequential recommendation models has been proposed before to provide superior recommendation results. Early works on sequential recommendation utilizes Markov chains \cite{shani2005mdp,rendle2010factorizing} to capture typical sequential patterns of user historical actions. With the advance of deep learning technologies, deep learning models, including RNN based, CNN based and attention based methods, achieve the state-of-the-art recommendation performance. 

RNN based models \cite{hidasi2015session,hidasi2018recurrent} leverage RNN or its variants, such as Gated Recurrent Unit (GRU)\cite{cho2014learning} and Long Short-Term Memory (LSTM)\cite{graves2012long}, to encode the user's previous actions into a compressed vector. With this vector as a context, these models are able to capture the dynamic and evolving interests of different users. Though many methods \cite{hidasi2015session,hidasi2018recurrent,yu2016dynamic,donkers2017sequential,li2017neural} with different loss functions and sampling strategies are proposed, they all belong to this category. For the problem of sequential recommendation, some adjacent actions may not have dependency relation (e.g. a user bought $i_3$ because of $i_0$ rather than $i_2$), which is inconsistent with the assumptions for model structure design of RNN. Moreover, because of the size limitation of the compressed vector, these models may have trouble to model long range impact among user actions.

Admittedly, CNN based architectures are not a natural way to capture sequential patterns. But Tang and Wang manage to propose a competitive Convolutional Sequence Model (Caser) \cite{tang2018personalized} to capture sequential patterns at point-level, union-level and skip behaviors with both horizontal and vertical convolutional filters. And NextItNet \cite{yuan2019simple} further improves the recommendation performance by addressing some design limitations of Caser.

As self-attention mechanism has shown promising results for sequential data \cite{vaswani2017attention,kenton2019bert}, we see rising enthusiasm to apply it to sequential recommendation. Kang et al. propose SASRec \cite{kang2018self} consisting of multi-layer transformer decoders to model user's sequential actions and achieve state-of-the-art performance on multiple datasets. Besides SASRec, with the introduction of Cloze task, BERT4Rec \cite{sun2019bert4rec} utilizes bidirectional self-attention networks to capture the pattern of user behaviors. Generally speaking, the self-attention based models can achieve supeior performance compared with the previous models.  
\vspace{-0.2cm}
\subsection{Existing Debias Methods for Popularity Bias}
Since the training data for recommender systems is collected from observational results rather than controlled experiments, the interaction data is missing-not-at-random and the popular items are over-represented in the training samples. Therefore, the recommender model will learn skewed user preference with the impact of popularity bias. As popularity bias is an important problem for recommender systems, it arouses rising interests from many researchers. To counteract the adverse effect of popularity bias, the existing methods can be roughly classified to three categories, namely post-processing with re-ranking, Inverse Propensity Weighting and Structural Causal models.
\vspace{-0.2cm}
\subsubsection{Re-Ranking}
These methods post-process the ranking model's results to balance the exposure rate of items from different popularity levels. Himan et al. \cite{abdollahpouri2019managing} increase the exposure of less popular items in recommendations with the application of a personalized diversification re-ranking approach. Moreover, Zhu \cite{zhu2021popularity} proposes an effective re-ranking methods to ensure fair exposure of items with equal user preference match. Nonetheless, these approaches usually promote diversity at the cost of the overall utility metrics. 
\vspace{-0.2cm}
\subsubsection{Inverse Propensity Weighting}
Multiple methods are proposed to learn unbiased models by utilizing inverse propensity weighting \cite{schnabel2016recommendations,swaminathan2015self,rosenbaum1983central}  to adjust the weight of different training samples. Rel-MF \cite{saito2020unbiased} derives the unbiased point-wise loss for enhanced recommendation qualtity. And MF-DU \cite{lee2021dual} achieves better bias correction by separately estimating the exposure probability for interacted and non-interacted data. For the pair-wise learning setting, Unbiased Bayesian Personalized Ranking \cite{saito2020unbiased2} extends the algorithm of BPR\cite{rendle2012bpr} and formulates an unbiased objective function. Recently, Wang et al. \cite{wang2022unbiased} proposes a bias correction approach for sequential recommendation. But to estimate the propensity with his approach, it can only work for the problem of predicting whether the user will positively rate the interacted items rather than the usual setting of estimating the probability of the next interaction item for sequential recommendation. These IPW related methods achieve success in many applications. However, their performance may be adversely affected by the high variance \cite{swaminathan2015self,bottou2013counterfactual,gruson2019offline} problem.  
\vspace{-0.2cm}
\subsubsection{Structural Causal Model} \label{sec:scm}
Structural causal model-based methods \cite{wei2021model,zheng2021disentangling,chen2022co,zhang2021causal,he2022addressing,wang2021deconfounded} achieve the state-of-the-art debiasing performance. By effectively modeling the confounding and colliding causal graph structures reflecting real-world relations, they are able to predict the direct and indirect effects of counterfactual interventions. DICE \cite{zheng2021disentangling} disentangles the user embedding to represent conformity and interest separately for causal recommendation. MACR \cite{wei2021model} proposes a model-agnostic counterfactual reasoning (MACR) framework and performs counterfactual inference to eliminate the impact of popularity bias. Compared with the previous two algorithms, PDA \cite{zhang2021causal} and $CD^2AN$ \cite{chen2022co} works in a slightly different setting by addressing the popularity drift issue (The item's popularity tends to vary with time). They estimate the item's future popularity and further leverage it to attain superior performance. In their setting, the data is separated based on different time slot. And they use the last slot's data rather than unbiased data for offline testing in their experiments.
\vspace{-0.2cm}
\section{Problem Definition}
For the sequential recommendation problem, there exist a set of users ($\mathcal{U} = \{u_i\}_{1 \leq i \leq N}$) and a candidate set of items ($\mathcal{I} = \{i_j\}_{1 \leq j \leq M}$). Moreover, each user $u$'s interaction history consists of a sequence of items from $\mathcal{I}$, $S^u=(s_1^u, s_2^u,...,s_{t-1}^u)$, where $s_j^u \in \mathcal{I}$. The index $j$ for $s_j^u$ denotes the step in which an action occurs in the interaction sequence. Given the current interaction history of $S^u$, sequential recommendation intends to predict the item that user $u$ will interact with at the next time step $t$. To attain the goal, the recommender model usually model the interaction probability over all possible items with point-wise loss function or estimate the relative order of different item pairs with pair-wise loss function \cite{burges2005learning,rendle2012bpr}. 

Popular items are exposed much more frequently than expected \cite{wei2021model}. Because of the skewed training data, all of the recommendation models are affected by the popularity bias \cite{chen2020bias} and may fail to recommend unpopular items, which match the user's interest better. Moreover, the popular items constitute the major part of the user's interaction sequence, which is an essential feature for sequential recommendation. Therefore, compared with other general recommendation models, the popularity bias \cite{chen2020bias} impacts the performance of sequential models more severely, which can be proven by Figure \ref{fig:rec_dist}. Consequently, it is highly desirable to apply effective debiasing methods to understand the user's preferences well for better recommendation performance. To assess the real recommendation quality, we follow the practice of prior works \cite{wei2021model,zheng2021disentangling,bonner2018causal} and perform unbiased evaluation with re-weighted testing data that is uniform distribution over items. 
\vspace{-0.2cm}
\section{Methodology}
In this section, we firstly present the causal graph and how to carry out counterfactual reasoning based on it for unbiased preference learning. Please note that this causal graph can apply to general recommendation models. In this paper, we concentrate on sequential models mainly because of their effectiveness and superior performance. Second, the general design of base sequential models are described in detail. Finally, we elaborate on how to apply the proposed causal graph to sequential recommendation by specifically augmenting the base sequential model structures and loss functions. For accurate bias correction, we design separate sequential structures based on the interest or popularity representations of historical interactions to precisely model the user's preference for interest or conformtiy. We also discuss why the proposed approach can effectively address the popularity bias issue.  

\begin{figure} 
  \centering 
  \subfigure[Real World]{ 
    \label{fig:causal:a} 
    \centering
    \includegraphics[width=1.5in]{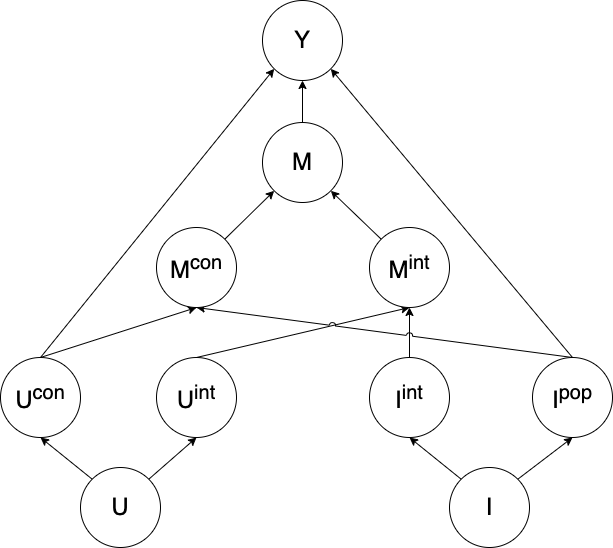} 
    }
    \subfigure[Counterfactual World]{ 
    \label{fig:causal:b} 
    \centering
    \includegraphics[width=1.5in]{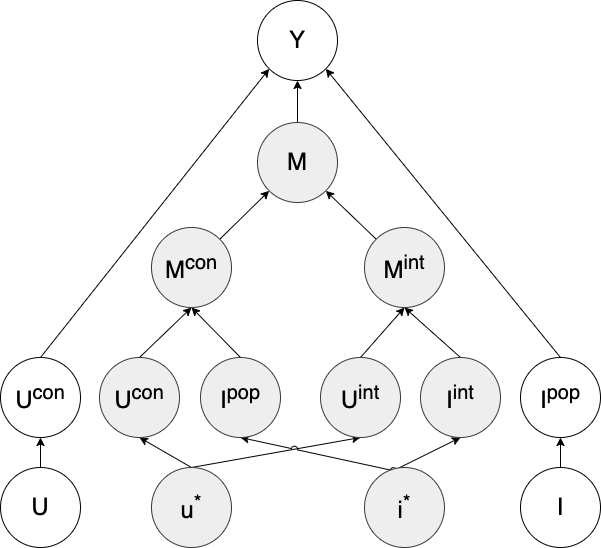} 
    }
    \caption{Comparison between real world and counterfactual
world causal graphs}
    \label{fig:causal}
\end{figure}

\begin{figure*}[!htbp]
\setlength{\abovecaptionskip}{-0.0cm} 
	\centering\textbf{}
	\includegraphics[scale=0.25]{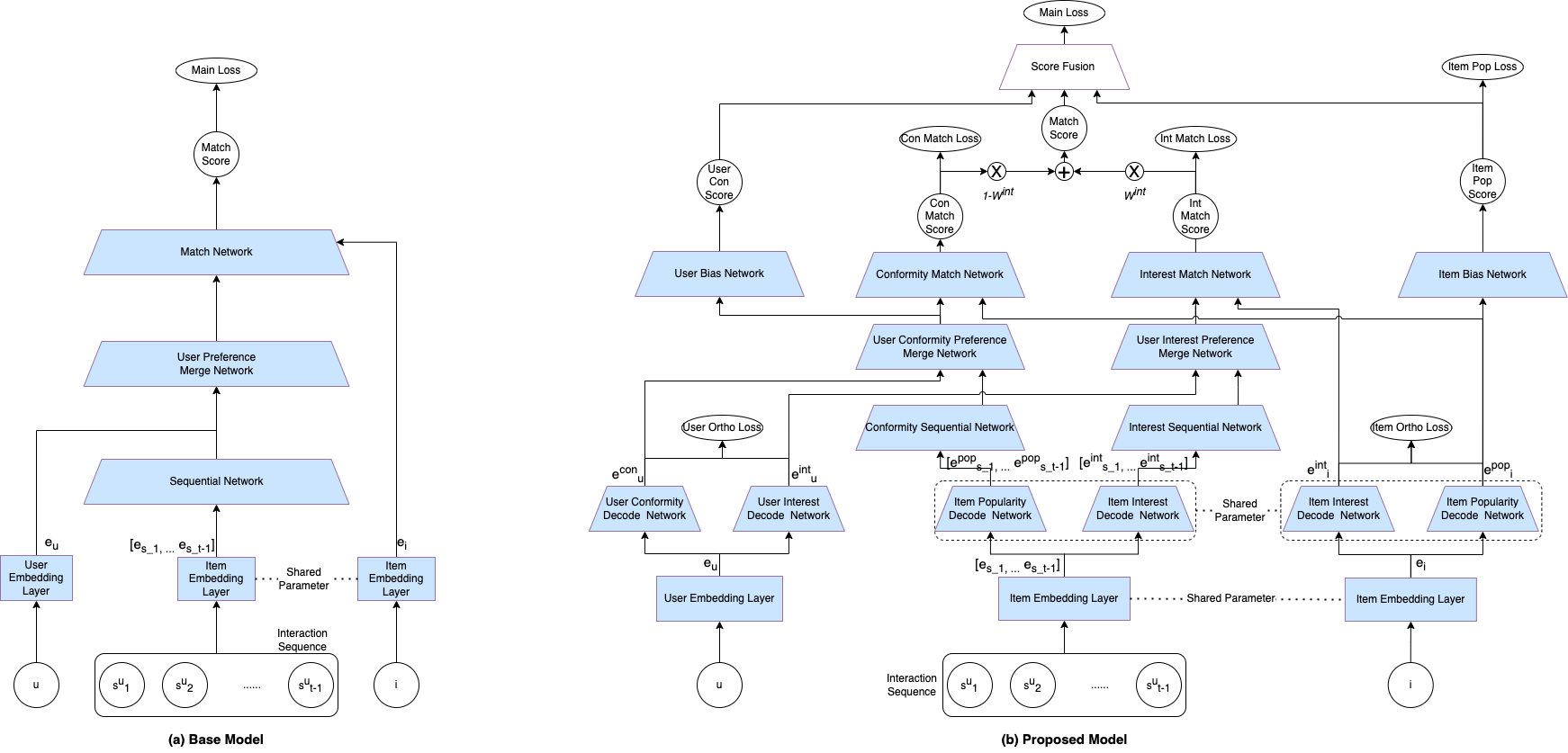}
	\caption{Model Architecture}
	\label{fig:model}
\end{figure*}
\vspace{-0.2cm}
\subsection{Causal Graph for Counterfactual Reasoning} \label{sec:causal_graph}
Each causal graph \cite{neuberg2003causality} is an instance of directed acyclic graph (DAG) $G=\{V,E\}$, where $V$ denotes the set of variables and $E$ represents the cause-effect relations among these variables. In the causal graph, the variables are represented with capital letters. And lowercase letter denotes an observed value for certain variable. There are both direct effects and indirect effects in the causal graph. For instance, in Figure \ref{fig:causal:a}, $I^{pop}$ exerts both direct effects (through the path of $I^{pop} \rightarrow Y$) and indirect effects (through $I^{pop} \rightarrow M^{con} \rightarrow M \rightarrow Y$) on $Y$. 

Please refer to Figure \ref{fig:causal:a} for the proposed causal graph. $U$ and $I$ denote the user and item embeddings respectively. First, for user embedding, we can further refine it to two representations to denote the user's preference for conformity and interest. In parallel, the item embedding can be refined to interest representations and popularity representations. Second, $M^{con}$ denotes the match between user conformity and item popularity. We use vectorized representations rather than scalar values to capture the diversity in user's varying conformity behaviors for different category items. Moreover, the match score $M$ is derived based on $M^{con}$ and $M^{int}$ (The interest match between user and item). The explicit modeling on the effect (interaction) and cause (conformity or interest) leads to robust models with stronger generalization capabilities \cite{zheng2021disentangling}, especially when the training and testing data are collected with different causes (conformity or interest). 

Besides the match score between the user and item, $I^{pop}$ and $U^{con}$ have direct impact on interaction $Y$. Popular items are exposed more by the existing recommender systems and interacted more by the users. And high conformity "easy" users tend to interact irrespective of his true preference. Unlike popularity and conformity, it is reasonable to suppose the interest representations only impact $Y$ through the match between the corresponding user and item. As item popularity directly impact $Y$, the model will assign unreasonable higher scores for popular items and recommend them more, thereby rendering the adverse feedback loop. In principal, removing the direct effect from item popularity and user conformity can help to eliminate the popularity bias, which means to mask the impact from the paths of $I \rightarrow I^{pop} \rightarrow Y$ and $U \rightarrow U^{con} \rightarrow Y$ 

Given a specific user item pair ($u, i$), let $U^{con}_u$, $I^{pop}_i$ and $M_{u,i}$ denote the conformity representation for $u$, popularity representation for $i$ and the overall match score for ($u,i$) pair. Then, the final prediction score can be represented with $Y_{U^{con}_u,I^{pop}_i,M_{u,i}}$. 

Reasoning in the counterfactual world (Figure \ref{fig:causal:b}) is necessary to estimate the unbiased user preference. In this counterfactual world, for the path of $U \rightarrow U^{con} \rightarrow Y$ and $I \rightarrow I^{pop} \rightarrow Y$, the previous user item pair ($u, i$) is still used. However, the match score $M_{u^*,i^*}$ is computed based on the pair of ($u^*,i^*$) where the reference status of $I=i^*$ and $U=u^*$ refer to the situation of masking $I$ and $U$ from real world as null or taking average values. Please note that $U$ and $I$ have different input values for varying paths, which can never happen in an real world scenario. In this counterfactual world, it is straightforward to derive the final prediction score as $Y_{U^{con}_u,I^{pop}_i,M_{u^*,i^*}}$. 

Accordingly, the unbiased preference can be obtained as follows:
\begin{fequation} \label{eq:tie}
unbiased\_score = Y_{U^{con}_u,I^{pop}_i,M_{u,i}} - Y_{U^{con}_u,I^{pop}_i,M_{u^*,i^*}}
\end{fequation}
which means the impact of $U$ and $I$ on $Y$ only through the paths of $I \rightarrow I^{int},I^{pop} \rightarrow M^{con},M^{int} \rightarrow M \rightarrow Y$ and $U \rightarrow U^{int},U^{con} \rightarrow M^{con},M^{int} \rightarrow M \rightarrow Y$. 
\vspace{-0.2cm}
\subsection{Base Sequential Models}
Please refer to Figure \ref{fig:model} (a) for the architecture of a general sequential recommendation model. These blocks in blue represent trainable components. Generally speaking, a typical sequential model takes three types of features as input, namely item features, user interaction sequence feature and other user features (e.g., user id and location). Usually, all of these features are high-dimensional binary features from one-hot encoding of categorical variables (e.g. user and item id) or discretization of dense variables (e.g., activity counting features). First, these features are projected to dense representations with the fully connected embedding layers.

\begin{fequation} \label{eq:u_embedding}
e_u = Embedding_u(u)
\end{fequation}
\vspace{-0.4cm}
\begin{fequation} \label{eq:i_embedding}
e_i = Embedding_i(i)
\end{fequation}
\vspace{-0.4cm}
\begin{fequation} \label{eq:u_sequence}
e_{s_1},e_{s_2},...,e_{s_{t-1}} = Embedding_{seq}(s_1^u, s_2^u,...,s_{t-1}^u)
\end{fequation}
where $Embedding_{seq}$ and $Embedding_i$ usually share the same parameters to boost the recommendation performance \cite{kang2018self}. And $e_u \in R^{1 \times d}$, $e_i \in R^{1 \times d}$ and $e_{s_j} \in R^{1 \times d}$ denote the $d$ dimensional user, item and sequence embeddings.

Then, a state-of-the-art sequential network, such as RNN\cite{hidasi2015session}, CNN\cite{yuan2019simple} or Transformer\cite{kang2018self}, accepts the sequence embedding $(e_{s1},e_{s2},...,e_{sk}$) and generates the user's dynamic preference.
\begin{fequation} \label{eq:seq}
pref^{dyn}_{u,t} = Seq(e_{s_1},e_{s_2},...,e_{s_{t-1}})
\end{fequation}
where $pref^{dyn}_{u,t}$ means the dynamic preference of $u$ at step $t$.

Moreover, to get accurate overall user preference, we can merge the information from $pref^{dyn}_{u,t}$ and $e_u$. Please note that this operation can involve a trainable network, a simple concatenation or just return $pref^{dyn}_{u,t}$ itself. Based on our experiments, we can achieve competitive results with $pref^{dyn}_{u,t}$ itself on the public datasets, which is consistent with the results of previous works \cite{kang2018self}. In this case, the explicit user embeddings are actually not used. For our industry data, we observe consistent gains by merging together these two representations of $pref^{dyn}_{u,t}$ and $e_u$.
\begin{fequation} \label{eq:merge_u}
pref_{u,t} = Merge(e_u, pref^{dyn}_{u,t})
\end{fequation}

Finally, with $pref_{u,t}$ and $e_i$ as inputs, the prediction score for step ($t$) can be estimated by the match network. This network can be a MLP or just dot product. 
\begin{fequation} \label{eq:match}
\hat{y}_{u,i,t} = Match(e_i, pref_{u,t})
\end{fequation}

For model training, the point-wise loss is calculated based on the match score and the corresponding training label. For pair-wise loss \cite{rendle2012bpr}, we need to compute the loss based on a user's match scores with different target items. 

For the results of equation \eqref{eq:u_embedding} to \eqref{eq:merge_u}, all of them are vectors. And the outcome of equation \eqref{eq:match} is a scalar value.
\vspace{-0.2cm}
\subsection{Proposed Model Architecture of DCR}
In this section, we describe the overall proposed model architecture. We will discuss how to counteract popularity bias with proper model training and inference in the following sections. For the results of equation \eqref{eq:Conformity_u} to \eqref{eq:pref_int}, all of the values are vectors. And the outcomes of equation \eqref{eq:match_con} to \eqref{eq:final_score} to are of scalar values.
\vspace{-0.2cm}
\subsubsection{Representation Disentanglement}
As illustrated in Figure \ref{fig:model} (b), the proposed model accepts the same set of features and leverages Equations \eqref{eq:u_embedding},\eqref{eq:i_embedding},\eqref{eq:u_sequence} to derive $e_u$, $e_i$ and the sequence embeddings of $e_{s_1},e_{s_2},...,e_{s_{t-1}}$. Then, we disentangle the embeddings of the target item and items in the user's historical interaction sequence with shared encoder networks to interest representations and popularity representations. In parallel, we disentangle the user embedding to interest representations and conformity representations. Since an overview of the whole picture is very helpful for understanding further design decisions, we delay the discussion of how to ensure high quality representation disentanglement with the introduction of well-designed training losses to the next sections. To represent item popularity and user conformity well, we do not think scalar values suffice. For example, the popularity level of items from different categories tend to have very different impact on interaction probabilities with the same user. As a result, we maintain vectorized popularity and conformity representations for fine-grained modeling. There operations can be formally defined below.
\begin{fequation} \label{eq:Conformity_u}
e^{con}_u = Conformity_u(e_u)
\end{fequation}
\vspace{-0.4cm}
\begin{fequation} \label{eq:Interest_u}
e^{int}_u = Interest_u(e_u)
\end{fequation}
\vspace{-0.4cm}
\begin{fequation} \label{eq:Popularity_i}
e^{pop}_i = Popularity_i(e_i)
\end{fequation}
\vspace{-0.4cm}
\begin{fequation} \label{eq:Interest_i}
e^{int}_i = Interest_i(e_i)
\end{fequation}
\vspace{-0.4cm}
\begin{fequation} \label{eq:Popularity_sequence}
e^{pop}_{s_j} = Popularity_{seq}(e_{s_j})_{1 \leq j \leq t-1}
\end{fequation}
\vspace{-0.4cm}
\begin{fequation} \label{eq:Interest_sequence}
e^{int}_{s_j} = Interest_{seq}(e_{s_j})_{1 \leq j \leq t-1}
\end{fequation}
where $Popularity_i$ and $Interest_i$ share the same parameters with $Popularity_{seq}$ and $Interest_{seq}$ correspondingly. And each equation denotes a trainable neural network module.

Please note that $e^{pop}_i$ and $e^{int}_i$ correspond to the implementation for the nodes of $I^{pop}$ and $I^{int}$ in Figure \ref{fig:causal:a} respectively.
\vspace{-0.2cm}
\subsubsection{User Preference Mining}
Then, separate sequential networks are leveraged to mine the user's dynamic preference for interest and conformity from the disentangled embeddings of the interaction item sequence. Specifically, we think the sequence of interest embeddings of historical interacted items can help to represent the user's interests. And we further argue that the user's conformity can be mined from $e^{pop}_{s_1},e^{pop}_{s_2},...,e^{pop}_{s_{t-1}}$. Additionally, we also merge with the disentangled representations from $e_u$ to derive the user's next step preference for interest and conformity. The results of equation \eqref{eq:pref_con} and \eqref{eq:pref_int}, namely $pref^{con}_{u,t}$ and $pref^{int}_{u,t}$,  stand for the implementation for the nodes of $U^{con}$ and $U^{int}$ in Figure \ref{fig:causal:a} respectively.
\begin{fequation} \label{eq:pref_con}
pref^{con}_{u,t} = Merge^{con}(Seq^{con}(e^{pop}_{s_1},e^{pop}_{s_2},...,e^{pop}_{s_{t-1}}),e^{con}_u) 
\end{fequation}
\vspace{-0.3cm}
\begin{fequation} \label{eq:pref_int}
pref^{int}_{u,t} = Merge^{int}(Seq^{int}(e^{int}_{s_1},e^{int}_{s_2},...,e^{int}_{s_{t-1}}),e^{int}_u) 
\end{fequation}
where $Seq^{con}$ and $Seq^{int}$ can be any sequential networks, including RNN\cite{hidasi2015session}, CNN\cite{yuan2019simple} and Self-Attention\cite{kang2018self}. And the merge operation can involve a trainable network, simple concatenation or just
emitting one of the inputs. 
\vspace{-0.2cm}
\subsubsection{User Item Matching}
At the next step, besides the interest match network, a standalone network is utilized to compute the match score between user conformity and item popularity. Moreover, we compute dynamic weights with MLP for the match scores of user interest and conformity with equation \eqref{eq:con_int_w} to capture their varying importance for different contexts. Finally, we can get the match score $\hat{y}_{m,u,i,t}$ for step $t$ with equation \eqref{eq:match_overall}.
\begin{fequation} \label{eq:match_con}
\hat{y}^{con}_{m,u,i,t} = Match^{con}(e^{pop}_i, pref^{con}_{u,t})
\end{fequation}
\vspace{-0.3cm}
\begin{fequation} \label{eq:match_int}
\hat{y}^{int}_{m,u,i,t} = Match^{int}(e^{int}_i, pref^{int}_{u,t})
\end{fequation}
\vspace{-0.3cm}
\begin{fequation} \label{eq:con_int_w}
w^{int} = \sigma(AttenNet(e^{int}_i, pref^{int}_{u,t}, e^{pop}_i, pref^{con}_{u,t}))
\end{fequation}
\vspace{-0.3cm}
\begin{fequation} \label{eq:match_overall}
\hat{y}_{m,u,i,t} = w^{int} * \hat{y}^{int}_{m,u,i,t} + (1.0 - w^{int}) * \hat{y}^{con}_{m,u,i,t}
\end{fequation}
where $\hat{y}^{con}_{m,u,i,t}$, $\hat{y}^{int}_{m,u,i,t}$ and $\hat{y}_{m,u,i,t}$ denote the implementation for the nodes of $M^{con}$, $M^{int}$ and $M$ in Figure \ref{fig:causal:a} respectively. And $\sigma$ stands for the sigmoid function. Moreover, the match function can involve a trainable MLP network or dot product operation.
\vspace{-0.2cm}
\subsubsection{Incorporating Direct Effect of User Conformity and Item Popularity}
Based on section \ref{sec:causal_graph}, besides the match score between user and item, we also need introduce the direct effect from user conformity and item popularity. First, we define influence from item popularity as below.
\begin{fequation} \label{eq:score_item}
\hat{y}_{i} = ItemPopularity(e^{pop}_i)
\end{fequation}
Furthermore, we represent the extent to which the user would interact with items regardless of preference match as follows.  
\begin{fequation} \label{eq:score_user}
\hat{y}_{u,t} = UserConformity(pref^{con}_{u,t})
\end{fequation}
Finally, these two scores are aggregated with the match score to acquire the final prediction score for step $k+1$ as below.
\begin{fequation} \label{eq:final_score}
\hat{y}_{u,i,t} = \hat{y}_{m,u,i,t} * \sigma(\hat{y}_{u,t}) * \sigma(\hat{y}_{i})
\end{fequation}
where $\hat{y}_{u,i,t}$ denotes the implementation for the node $Y$ in Figure \ref{fig:causal:a}. 

With this design, to generate high score for the interaction between a non-conformitive user and an unpopular item, the score of $\hat{y}_{m,u,i,t}$ must be pushed high during model training. 
\vspace{-0.2cm}
\subsection{DCR Model Training}
First, following \cite{wei2021model}, we take $\hat{y}_{u,i,t}$ from equation \eqref{eq:final_score} as prediction and define the main loss with binary cross entropy (BCE) loss \cite{xue2017deep} in equation \eqref{eq:main_loss}. 
\begin{fequation} \label{eq:main_loss}
L_{main} = -\displaystyle\sum_{(u,i,t) \in D} y_{u,i,t}  ln(\sigma(\hat{y}_{u,i,t}))+(1-y_{u,i,t})ln(1-\sigma(\hat{y}_{u,i,t}))
\end{fequation}
where $\sigma$ means the sigmoid function and $y_{u,i,t}$ is the ground truth label for step $t$. In our experiments, we will verify our proposed method also achieves substantial gains over the existing debiasing methods with pair-wise loss functions\cite{rendle2012bpr}.

Second, to make sure $e^{pop}_i$ contains the desired item popularity information, additional supervision is added for $\hat{y}_i$ from equation \eqref{eq:score_item}. As item popularity directly impacts interaction, we add the BCE loss for $\hat{y}_i$ below. 
\begin{fequation} \label{eq:i_loss}
L_{i} = -\displaystyle\sum_{(u,i,t) \in D} y_{u,i,t}  ln(\sigma(\hat{y}_{i}))+(1-y_{u,i,t})  ln(1-\sigma(\hat{y}_{i}))
\end{fequation}
Unlike MACR\cite{wei2021model}, we do not introduce the loss for the user tower as we find this loss has marginal impact on model performance, which is consistent with MACR's results.

Moreover, we also add supervision for $\hat{y}^{con}_{m,u,i,t}$ from equation \eqref{eq:match_con} and $\hat{y}^{int}_{m,u,i,t}$ from \eqref{eq:match_int}. As $\hat{y}^{int}_{m,u,i,t}$ should capture the user's unbiased interests, we leverage the proposed BCE based IPW loss proposed in \cite{lee2021dual}, which achieves superior performance among the point-wise IPW methods. Specifically, they propose to define propensity scores for positive and negative items separately as below.  
\begin{fequation}
\theta_{*,i}^{+}=(\frac{n_i}{max_{i \in I}n_i})^{\omega} \;\;\;\; \theta_{*,i}^{-}=(1-\frac{n_i}{max_{i \in I}n_i})^{\rho}
\end{fequation}
where $n_i$ signifies the number of interactions to the item $i$ by all users. And $\omega,\rho \in [0,1]$, which are both empirically set to 0.5 in their work, are the hyperparameters to control the skewness of item popularity. Then, we can define the supervision for $\hat{y}^{int}_{m,u,i,t}$ as below. Though IPW loss has the large variance issue, the impact here is much lighter as the auxiliary loss usually uses a much smaller weight than the main loss.
\begin{fequation}\label{eq:int_ipw_loss}
L^{int}_{m} = -\displaystyle\sum_{(u,i,t) \in D} \frac{y_{u,i,t}}{\theta_{*,i}^{+}}ln(\sigma(\hat{y}^{int}_{m,u,i,t}))+\frac{(1-y_{u,i,t})}{\theta_{*,i}^{-}}ln(1-\sigma(\hat{y}^{int}_{m,u,i,t}))
\end{fequation}
As $\hat{y}^{con}_{m,u,i,t}$ should also be able to explain the existing interactions. Similar to $\hat{y}_i$, we add the BCE loss below. With $L^{con}_{m}$, in order to match well with $e^{pop}_i$, $pref^{con}_{u,t}$ should approximate to the user's conformity. 
\begin{fequation}\label{eq:pop_bce_loss} 
L^{con}_{m} = -\displaystyle\sum_{(u,i,t) \in D} y_{u,i,t}  ln(\sigma(\hat{y}^{con}_{m,u,i,t}))+(1-y_{u,i,t})  ln(1-\sigma(\hat{y}^{con}_{m,u,i,t}))
\end{fequation}

$pref^{int}_{u,t}$ from \eqref{eq:pref_int}, $e^{int}_u$ and $e^{int}_i$ should approximate to the unbiased interest representations owing to the IPW loss of $L^{int}_m$. Moreover, with the aforementioned auxiliary loss of $L_i$ and $L^{con}_m$, $pref^{con}_{u,t}$ from \eqref{eq:pref_con}, $e^{con}_u$ and $e^{pop}_i$ are forced to contain conformity and popularity information. But $pref^{con}_{u,t}$, $e^{con}_u$ and $e^{pop}_i$ may still undesirably contain some interest information in theory. For proper representation disentanglement, $e^{con}_u$ and $e^{int}_u$ are enforced to be  independent with orthogonality loss. And the same loss is applied to $e^{pop}_i$ and $e^{int}_i$. We list the two orthogonality loss as below.
\begin{fequation}\label{eq:u_ortho_loss}
L^{ortho}_u = \left|CosineSimilarity(e^{con}_u, e^{int}_u)\right|
\end{fequation}
\vspace{-0.4cm}
\begin{fequation}\label{eq:i_ortho_loss}
L^{ortho}_i = \left|CosineSimilarity(e^{pop}_i, e^{int}_i)\right|
\end{fequation}

Finally, we can leverage the multi-task learning schema and use the loss below to optimize the proposed model.
\begin{fequation}\label{eq:total_loss}
L = L_{main} + \alpha(L^{int}_{m} + L^{con}_{m}) + \beta L_i + \gamma (L^{ortho}_u + L^{ortho}_i)
\end{fequation}
where $\alpha$, $\beta$ and $\gamma$ are the hyper-parameters to be tuned. We will carry out well-designed sensitivity tests to investigate their impact.
\vspace{-0.2cm}
\subsection{DCR Model Inference}
As aforementioned in section \ref{sec:causal_graph}, the popularity bias can be eliminated if we can remove the effect of the path $U -> U^{con} -> Y$ and $I -> I^{pop} -> Y$. For unbiased recommendation, we propose to adjust the score $\hat{y}_{u,i,t}$ from equation \eqref{eq:final_score} for ranking as below:
\begin{fequation} 
\hat{y}_{u,i,t} - c * \sigma(\hat{y}_{u,t}) * \sigma(\hat{y}_{i})
\end{fequation}
where $c$ represents the reference status of $\hat{y}_{m,u,i,t}$ from eq \eqref{eq:match_overall}, namely the $M$ node in Figure \ref{fig:causal:b}. 

The above inference is an exact implementation for equation \eqref{eq:tie}. $\hat{y}_{u,i,t}$ denotes the score from the real world while $c * \sigma(\hat{y}_{u,t}) * \sigma(\hat{y}_{i})$ acts as the score from the counterfactual world. With the help of counterfactual reasoning and disentangled representations, the recommendation model can attain superior recommendation results.

\vspace{-0.2cm}
\section{Experiments}
\begin{table}
\normalfont
\centering
\caption{The dataset statistics for user count, item count, interaction count and gini index \cite{enwiki:1101796604} computed from item occurrence frequency.}
\label{tab:data_statistics}
\begin{tabular}{lc|c|c} 
\hline
                  & MovieLens-1M & Video Games & Steam    \\ 
\hline\hline
User Count        & 6,040         & 31,013       & 334,730   \\
Item Count        & 3,416         & 23,715       & 13,047    \\
Interaction Count & 999,611       & 287,107      & 3,686,172  \\
Item Gini Index   & 0.6036       & 0.6567      & 0.8398   \\
\hline\hline

\end{tabular}
\end{table}

\begin{table*}[ht]
\footnotesize
\centering
\caption{The performance evaluation of the compared methods with NDCG@10 and Hit Rate@10. The bold-face font denotes the winner in that row. The best results with "*" symbol indicate p < 0.05 for one-tailed t-test with the best competitor. Moreover, the underline symbol denotes the best results excluding DCR. }
\label{tab:overview}
\begin{tabular}{lllcc|cccccc|llll|c} 
\hline
SeqNet                     & Data                   & Metric  & $Base_{bce}$ & $Base_{bpr}$ & $\makecell[c]{Bias\\Tower}$ & $IPW_{bce}$ & $IPW_{bpr}$      & $DICE$         & $PD$ & $MACR$         & $\makecell[c]{DCR\\var0}$ & $\makecell[c]{DCR\\var1}$ & $\makecell[c]{DCR\\var2}$ & $DCR$           & $RelGain$  \\ 
\hline\hline
\multirow{6}{*}{SASRec\cite{kang2018self}}    & \multirow{2}{*}{ML-1M} & NDCG    & 0.3732     & 0.3704     & 0.3993                      & 0.3931    & 0.3946         & 0.3858         & 0.3842         & \uline{0.4077} & 0.4137                    & 0.4171                    & 0.4325                    & $\boldsymbol{0.4438}^*$ & 8.85\%     \\
                           &                        & HitRate & 0.5725     & 0.5715     & 0.6011                      & 0.5911    & 0.5902         & 0.5888         & 0.5830         & \uline{0.6208} & 0.6279                    & 0.6207                    & 0.6427                    & $\boldsymbol{0.6680}^*$ & 7.60\%     \\ 
\cline{2-16}
                           & \multirow{2}{*}{Games} & NDCG    & 0.2962     & 0.3088     & 0.3054                      & 0.2971    & \uline{0.3175} & 0.2968         & 0.3145         & 0.2954         & 0.3022                    & 0.3081                    & 0.3174                    & $\boldsymbol{0.3217}^*$ & 1.32\%     \\
                           &                        & HitRate & 0.4693     & 0.4961     & 0.4794                      & 0.4717    & \uline{0.5019} & 0.4894         & 0.4856         & 0.4692         & 0.4789                    & 0.4808                    & 0.4979                    & $\boldsymbol{0.5059}$   & 0.80\%     \\ 
\cline{2-16}
                           & \multirow{2}{*}{Steam} & NDCG    & 0.1989     & 0.1927     & 0.2133                      & 0.2023    & 0.2138         & \uline{0.2369} & 0.2142         & 0.2216         & 0.2297                    & 0.2402                    & 0.2441                    & $\boldsymbol{0.2556}^*$ & 7.89\%     \\
                           &                        & HitRate & 0.3446     & 0.3389     & 0.3683                      & 0.3453    & 0.3597         & \uline{0.3946} & 0.3642         & 0.3723         & 0.3995                    & 0.3978                    & 0.4057                    & $\boldsymbol{0.4304}^*$ & 9.07\%     \\ 
\hline\hline
\multirow{6}{*}{GRU4Rec\cite{hidasi2015session}}   & \multirow{2}{*}{ML-1M} & NDCG    & 0.3586     & 0.3565     & 0.3783                      & 0.3749    & 0.3797         & 0.3738         & 0.3763         & \uline{0.3878} & 0.4043                    & 0.4078                    & 0.4190                    & $\boldsymbol{0.4346}^*$ & 12.07\%    \\
                           &                        & HitRate & 0.5525     & 0.5501     & 0.5724                      & 0.5685    & 0.5730         & 0.5668         & 0.5717         & \uline{0.5984} & 0.6135                    & 0.6095                    & 0.6263                    & $\boldsymbol{0.6487}^*$ & 8.41\%     \\ 
\cline{2-16}
                           & \multirow{2}{*}{Games} & NDCG    & 0.2315     & 0.2529     & 0.2289                      & 0.2292    & \uline{0.2597} & 0.2413         & 0.2533         & 0.2389         & 0.2264                    & 0.2448                    & 0.2575                    & $\boldsymbol{0.2636}^*$ & 1.50\%     \\
                           &                        & HitRate & 0.3860     & 0.4259     & 0.3924                      & 0.3854    & \uline{0.4292} & 0.4169         & 0.4148         & 0.4006         & 0.3832                    & 0.4012                    & 0.4279                    & $\boldsymbol{0.4337}$   & 1.05\%     \\ 
\cline{2-16}
                           & \multirow{2}{*}{Steam} & NDCG    & 0.1640     & 0.1637     & 0.1661                      & 0.1554    & 0.1784         & \uline{0.2099} & 0.1779         & 0.1769         & 0.1918                    & 0.2057                    & 0.2078                    & $\boldsymbol{0.2210}^*$ & 5.29\%     \\
                           &                        & HitRate & 0.2917     & 0.2966     & 0.2969                      & 0.2734    & 0.3103         & \uline{0.3560} & 0.3169         & 0.3096         & 0.3426                    & 0.3446                    & 0.3487                    & $\boldsymbol{0.3758}^*$ & 5.56\%     \\ 
\hline\hline
\multirow{6}{*}{NextItNet\cite{yuan2019simple}} & \multirow{2}{*}{ML-1M} & NDCG    & 0.3560     & 0.3469     & 0.3778                      & 0.3673    & 0.3748         & 0.3739         & 0.3637         & \uline{0.3855} & 0.3641                    & 0.3922                    & 0.3953                    & $\boldsymbol{0.4124}^*$ & 6.98\%     \\
                           &                        & HitRate & 0.5463     & 0.5421     & 0.5706                      & 0.5651    & 0.5687         & 0.5683         & 0.5509         & \uline{0.6002} & 0.5732                    & 0.5902                    & 0.6097                    & $\boldsymbol{0.6317}^*$ & 5.25\%     \\ 
\cline{2-16}
                           & \multirow{2}{*}{Games} & NDCG    & 0.2413     & 0.2580     & 0.2498                      & 0.2306    & 0.2682         & 0.2440         & \uline{0.2697} & 0.2421         & 0.2496                    & 0.2684                    & 0.2828                    & $\boldsymbol{0.2896}^*$ & 7.38\%     \\
                           &                        & HitRate & 0.3993     & 0.4341     & 0.4175                      & 0.3902    & \uline{0.4456} & 0.4241         & 0.4308         & 0.4002         & 0.4211                    & 0.4364                    & 0.4605                    & $\boldsymbol{0.4636}^*$ & 4.04\%     \\ 
\cline{2-16}
                           & \multirow{2}{*}{Steam} & NDCG    & 0.1686     & 0.1625     & 0.1743                      & 0.1618    & 0.1759         & \uline{0.2047} & 0.1792         & 0.1795         & 0.1877                    & 0.2067                    & 0.2064                    & $\boldsymbol{0.2162}^*$ & 5.62\%     \\
                           &                        & HitRate & 0.2979     & 0.2927     & 0.3049                      & 0.2835    & 0.3043         & \uline{0.3463} & 0.3164         & 0.3152         & 0.3363                    & 0.3479                    & 0.3562                    & $\boldsymbol{0.3744}^*$ & 8.11\%     \\
\hline\hline

\end{tabular}

\end{table*}

\begin{table*}
\normalfont
\centering
\caption{The performance evaluation of tuning $\alpha$ and $\beta$ on MovieLens-1M}
\label{tab:sensitivity_test}
\begin{tabular}{lccc|cc|cc|cc|cc} 
\hline
\multirow{2}{*}{Backbone} & \multirow{2}{*}{\makecell[c]{Loss Weight\\ to tune}} & \multicolumn{2}{c|}{2e-4} & \multicolumn{2}{c|}{2e-3} & \multicolumn{2}{c|}{2e-2}         & \multicolumn{2}{c|}{2e-1} & \multicolumn{2}{c}{2e0}  \\ 
\cline{3-12}
                          &                                 & NDCG   & Hit Rate         & NDCG   & Hit Rate         & NDCG            & Hit Rate        & NDCG   & Hit Rate         & NDCG   & Hit Rate        \\ 
\hline\hline
\multirow{2}{*}{SASRec\cite{kang2018self}}   
                          & $\alpha$                        & 0.4265 & 0.6319           & 0.4223 & 0.6317           & \textbf{0.4438} & \textbf{0.6680} & 0.4155 & 0.6565           & 0.4117 & 0.6497          \\ 
                          & $\beta$                         & 0.3168 & 0.5128           & 0.3570 & 0.5617           & \textbf{0.4438} & \textbf{0.6680} & 0.3877 & 0.6221           & 0.3086 & 0.5076          \\                          
\hline\hline
\multirow{2}{*}{GRU4Rec\cite{hidasi2015session}} 
                          & $\alpha$                        & 0.4049 & 0.6209           & 0.4073 & 0.6113           & \textbf{0.4346} & \textbf{0.6487} & 0.4180 & 0.6431           & 0.4015 & 0.6264          \\
                          & $\beta$                         & 0.2692 & 0.4530           & 0.3324 & 0.5316           & \textbf{0.4346} & \textbf{0.6487} & 0.3698 & 0.5899           & 0.3101 & 0.5023          \\                          
\hline\hline
\end{tabular}
\end{table*}
In this section, we conduct extensive experiments with the aim of answering the following questions:

\begin{itemize}
\item[\textbf{RQ1}] Compared with the existing debiasing methods, can DCR achieve significant gains for a variety of base sequential models on multiple datasets ?
\item[\textbf{RQ2}] Can we verify the effectiveness and necessity of the introduced individual components?
\item[\textbf{RQ3}] How sensitive do the multiple hyper-parameters, including $\alpha$, $\beta$, $\gamma$ and $c$, impact the overall recommendation quality?
\item[\textbf{RQ4}] Can DCR achieve online gains over the existing state of the art debiasing methods if deployed on industrial recommendation service ?
\end{itemize}
\vspace{-0.2cm}
\subsection{Offline Experimental Settings}
\subsubsection{Datasets} 
We evaluate our methods on three public benchmark datasets, whose statistics are shown in Table \ref{tab:data_statistics}. 
\vspace{-0.07cm}
\begin{itemize}[leftmargin=.2in]
\item \textbf{MovieLens-1M}\cite{harper2015movielens}: One of the currently released MovieLens datasets, which contains 1 million movie ratings from about six thousand users across more than 3,000 movies.
\item \textbf{Amazon}\cite{mcauley2015image}: A series of datasets consisting of product reviews from Amazon.com. We use the sub-category of "Video Games".
\item \textbf{Steam}\cite{kang2018self}: A dataset crawled from an online game distribution platform named Steam. It includes game reviews and some other useful information, such as developer, category and pricing.
\end{itemize}
From the gini index \cite{enwiki:1101796604} computed based on item occurrence frequency, we can apparently observe the existence of popularity bias in these datasets. We follow the same pre-processing procedure as \cite{kang2018self} and discard rare users and items with lower than 5 related interactions. For each user, we treat the presence of a review or rating as implicit feedback and sort chronologically to decide the sequence order. For data partitioning, we treat the most recent action of each user ($S^u_{|S^u|}$) as testing. And the second most recent action ($S^u_{|S^u|-1}$) is left for validation.  All the previous items at $S^u$ are for training.
\vspace{-0.2cm}
\subsubsection{Experimental Models}
To verify the effectiveness, we conduct experiments on three popular sequential networks, including RNN based\cite{hidasi2015session}, CNN based\cite{yuan2019simple} and Self-Attention based\cite{kang2018self}. Based on our tests, adding explicit user embeddings does not achieve significant performance gains for the public data, which is consistent with \cite{kang2018self}. Thus, by default, the user preference merge networks at Figure \ref{fig:model} just copy the results of the sequential networks. But with specific experiment on the public data and the industrial model deployed for online A/B testing, DCR shows good performance for the configuration with explicit user embeddings. For the match network in Figure \ref{fig:model}, we use simple dot product as we do not observe gains with complex MLP networks. For the embedding disentanglement networks, we use two layer MLP structure of (150, 50) for interest while (100, 50) for popularity and conformity. We fix the item embedding's size to 50. For GRU4Rec\cite{hidasi2015session}, a two layer GRU networks with hidden size 50 is used. For NextItNet\cite{yuan2019simple}, we set the channel number to 50, kernel size to 3, and utilize the 8 layers convolutions with the dilation configuration of \{1,2,4,8,1,2,4,8\} proposed in the paper. For SASRec\cite{kang2018self}, we use two layer transformer networks \cite{vaswani2017attention} with hidden size of 50. Following \cite{kang2018self}, we set the dropout rate to 0.2 for MovieLens-1M while 0.5 for other sparser datasets. We tune the hyperparameters using the validation set and terminate training if its performance does not improve for 40 epochs. 

For each backbone sequential network, we test the following methods and run 10 times to report the results. For DCR and its variants, we use the architecture depicted at Figure \ref{fig:model}(b). For other methods, we use the architecture of Figure \ref{fig:model}(a) and take $pref_{u,t}$ from Eq \eqref{eq:merge_u} as the user embedding for further processing.
\vspace{-0.1cm}
\begin{itemize}[leftmargin=.2in]
\item \textbf{$Base_{bce}$}: This is the base method with BCE\cite{xue2017deep} loss.
\item \textbf{$Base_{bpr}$}: This is the base method with BPR\cite{rendle2012bpr} loss.
\item \textbf{$BiasTower$}\cite{koren2009matrix}: With BCE loss, it learns a bias score with the item tower, which is added to the prediction during training while removed during inference.
\item \textbf{$IPW_{bce}$}\cite{lee2021dual}: A SOTA IPW debiasing method with BCE loss.
\item \textbf{$IPW_{bpr}$}\cite{saito2020unbiased2}: A SOTA IPW debiasing method with BPR loss.
\item \textbf{$DICE$}\cite{zheng2021disentangling}: A structural causal model method that disentangles conformity and interest embeddings for superior performance.
\item \textbf{$PD$}\cite{zhang2021causal}: A structural causal model method that performs deconfounded training with do-calculus. As we care unbiased prediction rather than popularity drift here, PD is tested here instead of PDA.
\item \textbf{$MACR$}\cite{wei2021model}: A structural causal model method conducting counterfactual reasoning for unbiased prediction.
\item \textbf{$DCR_{var0}$}: $w^{int}$ in Equation \eqref{eq:con_int_w} is fixed to 1.0. This actually masks the match between user conformity and item popularity embeddings to show the impact.
\item \textbf{$DCR_{var1}$}: We do not use Equation \eqref{eq:con_int_w} and directly add $\hat{y}^{con}_m$ and $\hat{y}^{int}_m$ to compute $\hat{y}_m$. And $\alpha$ in Equation \eqref{eq:total_loss} is set to 0.0 to totally disable $L^{int}_m$ and $L^{con}_m$. 
\item \textbf{$DCR_{var2}$}: We do not use Equation \eqref{eq:con_int_w} and directly add $\hat{y}^{con}_m$ and $\hat{y}^{int}_m$ to compute $\hat{y}_m$. 
\item \textbf{$DCR$}: The proposed approach.
\end{itemize}
For DCR, we tune $\alpha$ and $\beta$ from \{2e-4,2e-3,2e-2,2e-1,2e0\}. DCR's $\gamma$ is tuned from 5e-4 to 50. As for $c$ of DCR and MACR, it is tuned from 0 to 80 with a step of 10. Following \cite{zhang2021causal}, for PD, we tune its hyperparameter of $\gamma$ from 0.02 to 0.25. For DICE, we also carefully tune its curriculum learning hyper-parameters following \cite{zheng2021disentangling}.
\vspace{-0.2cm}
\subsubsection{Evaluation Metrics}
We adopt two common metrics for recommendation, namely NDCG@10 and Hit Rate@10 \cite{he2017neural}. Hit Rate counts the fraction of times that the ground-truth
next item is among the top K items, while NDCG assigns heavier weight for front positions. To alleviate the heavy computation, following \cite{kang2018self,he2017neural}, we randomly sample 100 negative items for each ground-truth item and rank these items together to report the result. Moreover, for unbiased evaluation, we re-weight each testing instance with Inverse Propensity Weighting of the positive item to compute the mean value of NDCG and Hit Rate. We also calculate NDCG and Hit Rate with K=5 and 20 and find pretty consistent trend, which are omitted for space constraint. 

\subsection{Overall Offline Performance(\textbf{RQ1})}
Please refer to table \ref{tab:overview} for the overall results. Except for the hit rate metric of SASRec and GRU4Rec model on the Video Games dataset, DCR achieves significant relative gains over the best competitors. Among the competitors, $MACR$, $DICE$, $PD$ and $IPW_{bpr}$ performs the best for different scenarios. Comparing $IPW_{bpr}$ and $IPW_{bce}$, we can see $IPW_{bpr}$ generates better results. And $IPW_{bce}$ is sometimes worse than $Base_{bce}$, which also reflects the large variance issue of IPW methods. Comparing $MACR$ and $BiasTower$, we can see $BiasTower$ performs relatively stable across different datasets while $MACR$'s results fluctuate. We further investigate why $MACR$ performs bad on video games dataset and find its $L_i$ in equation \eqref{eq:i_loss} is much worse than $DCR$ in those scenarios. We think the reason is that $MACR$'s representations are affected more by the potential negative transfer among multiple losses without embedding disentanglement. $DICE$ performs well for Steam while not so satisfactory for the other two datasets, for which we cannot tune well for the curriculum learning hyper-parameters of $DICE$.  

\vspace{-0.2cm}
\subsection{Further Analysis for Offline Experiments(\textbf{RQ2\&RQ3})}
\subsubsection{Ablation Studies}
Please refer to table \ref{tab:overview}, the three DCR variants perform worse than DCR, which proves the necessity of introducing the corresponding components. Comparing with others, $DCR_{var0}$ performs worst indicating the necessity to consider the match between user conformity and item popularity. Furthermore, $DCR_{var1}$ performs worse than $DCR_{var2}$, which verifies $L^{int}_{m}$ at equation \eqref{eq:int_ipw_loss} and $L^{con}_{m}$ at \eqref{eq:pop_bce_loss} can help to boost the recommendation performance. Finally, comparing $DCR_{var2}$ and $DCR$, we are sure the attention score introduced in Equation \eqref{eq:con_int_w} for conformity match and interest match is essential for superior results.
\vspace{-0.2cm}
\subsubsection{Sensitivity Tests}
For the hyperparameters of multi-task loss weights, please refer to table \ref{tab:sensitivity_test} for the model performance with varying $\alpha$ and $\beta$ and refer to table \ref{tab:ortho_test} for tuning $\gamma$. Firstly, tuning these parameters will impact the recommendation quality, which also shows the effectiveness of the corresponding components. Moreover, the performance change gradually and predictably. And the best results for different datasets and models can be obtained with the same configuration. Among the three, the model is most sensitive to the change of $\beta$. For $\gamma$, initially, the model will perform better with the effective improvement of orthogonality loss from Equation \eqref{eq:i_ortho_loss}. But once the orthogonality loss is near 1e-3, further increasing $\gamma$ will produce worse results since the main loss can be negatively impacted with a too large $\gamma$. 

Please refer to table \ref{tab:c_test} for the impact of $c$. We can see tuning $c$ well is important for the model performance. And for different datasets, the best $c$ can vary greatly but the performance trend is predictable. Overall, DCR can be practically tuned and applied well.
\vspace{-0.2cm}

\begin{figure} [!ht]
  \centering 
  \subfigure[$Base_{bce}$'s NDCG over Emb Dim]{ 
    \label{fig:explicit_u_emb:a} 
    \centering
    \includegraphics[width=1.5in]{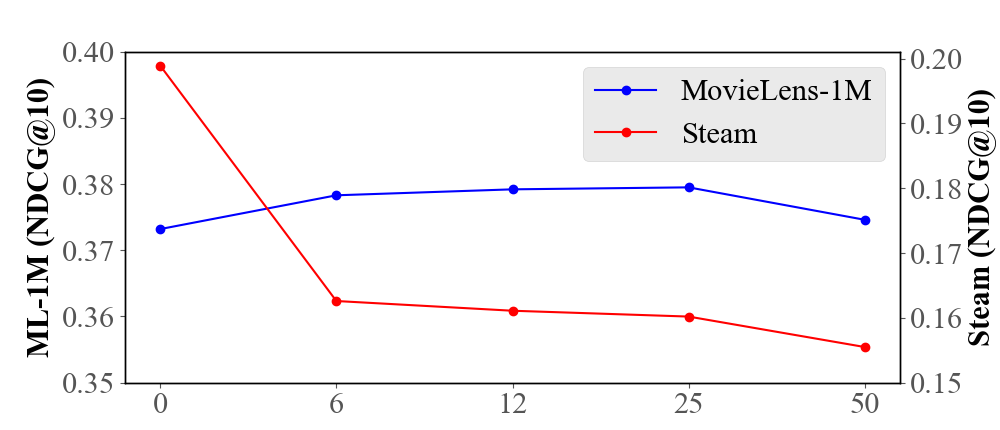} 
    }
    \subfigure[$DCR$ vs $Base_{bce}$]{ 
    \label{fig:explicit_u_emb:b} 
    \centering
    \includegraphics[width=1.6in]{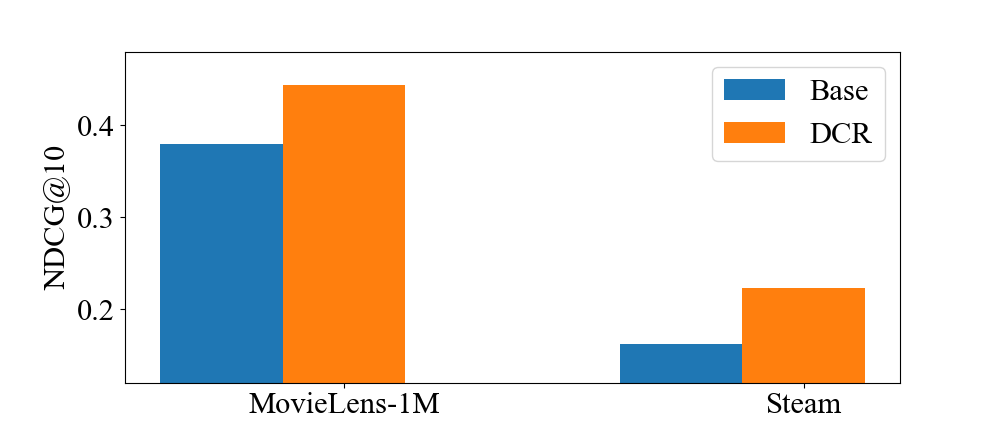} 
    }
    \caption{SASRec Results with Explicit User Embeddings}
    \label{fig:explicit_u_emb}
\end{figure}
\vspace{-0.2cm}
\subsubsection{The Effect of Explicit User Embedding}
We also test SASRec's performance of adding explicit user embeddings with different dimensions, disentangling them to conformity and interest representations and leveraging the two layer MLP structure (50,50) to merge with the sequential embeddings. We see relatively marginal gains for MovieLens-1M and Video Games while much worse results for Steam. For Steam, there are so many users so adding user embedding may cause the overfitting issue. We plot $Base_{bce}$'s NDCG@10 results of MovieLens-1M and Steam at Figure \ref{fig:explicit_u_emb:a}.  Moreover, excluding size 0, MovieLens-1M and Steam's best dimension size are 25 and 6 respectively. Based on such settings, DCR still achieves significant NDCG gains according to Figure \ref{fig:explicit_u_emb:b}.
\vspace{-0.2cm}

\begin{table}
\small
\centering
\caption{The effect of tuning $\gamma$ on the backbone of SASRec}
\label{tab:ortho_test}
\begin{tabular}{lcccc} 
\hline
Dataset                             & $\gamma$ & NDCG@10 & Hit Rate@10 & Orthogonality  \\ 
\hline
\multirow{7}{*}{\makecell[c]{MovieLens\\1M}}       & 0        & 0.4317  & 0.6451      & 0.155          \\
                                    & 5e-4     & 0.4363  & 0.6579      & 0.099          \\
                                    & 5e-3     & 0.4387  & 0.6495      & 0.061          \\
                                    & 5e-2     & 0.4378  & 0.6669      & 0.016          \\
                                    & 5e-1     & \textbf{0.4438}  & \textbf{0.6680}      & 0.002          \\
                                    & 5e0      & 0.4216  & 0.6427      & 0.001          \\
                                    & 5e1      & 0.4011  & 0.6322      & 0.001          \\ 
\hline
\multirow{7}{*}{\makecell[c]{Amazon \\Video Games}} & 0        & 0.3091  & 0.4982      & 0.153          \\
                                    & 5e-4     & 0.3128  & 0.4962      & 0.051          \\
                                    & 5e-3     & 0.3136  & 0.4993      & 0.027          \\
                                    & 5e-2     & 0.3135  & 0.5052      & 0.014          \\
                                    & 5e-1     & \textbf{0.3217}  & \textbf{0.5059}      & 0.001          \\
                                    & 5e0      & 0.3118  & 0.4931      & 0.0005         \\
                                    & 5e1      & 0.2955  & 0.4818      & 0.0001         \\
\hline
\end{tabular}
\end{table}

\begin{table}
\small
\centering
\caption{The effect of tuning $c$ on the backbone of SASRec}
\label{tab:c_test}
\begin{tabular}{lccc} 
\hline
Dataset                             & $c$ & NDCG@10         & Hit Rate@10      \\ 
\hline
\multirow{9}{*}{MovieLens-1M}       & 0   & 0.3866          & 0.5902           \\
                                    & 10  & 0.4223          & 0.6361           \\
                                    & 20  & 0.4306          & 0.6480           \\
                                    & 30  & \textbf{0.4438} & \textbf{0.6680}  \\
                                    & 40  & 0.4273          & 0.6604           \\
                                    & 50  & 0.4239          & 0.6615           \\
                                    & 60  & 0.3903          & 0.6183           \\
                                    & 70  & 0.3759          & 0.6096           \\
                                    & 80  & 0.3662          & 0.5952           \\ 
\hline
\multirow{9}{*}{Amazon Video Games} & 0   & 0.2903          & 0.4699           \\
                                    & 10  & 0.2909          & 0.4741           \\
                                    & 20  & 0.2979          & 0.4803           \\
                                    & 30  & 0.3029          & 0.4870           \\
                                    & 40  & 0.3006          & 0.4855           \\
                                    & 50  & 0.3132          & 0.4958           \\
                                    & 60  & 0.3159          & 0.5020           \\
                                    & 70  & \textbf{0.3217} & \textbf{0.5059}  \\
                                    & 80  & 0.3096          & 0.4979           \\
\hline
\end{tabular}
\end{table}

\subsection{Online A/B Testing(\textbf{RQ4})}
We applied the proposed method to the ranking stage \cite{covington2016deep} of an industrial large-scale video recommender system. As the sequential feature of previous user actions is the most important feature for the ranking model based on our analysis, it is a typical sequential recommender system. In contrast to the aforementioned public data listed in table \ref{tab:data_statistics}, other user features, such as user id, also generate considerable gains for the industrial data. The main online metrics include effective VV(count of Video Views with watch time exceeding a threshold) and total video watch time. Initially, we applied MACR\cite{wei2021model} and saw a significant (p<0.05) increase of 0.79\% for effective VV while neutral result for video watch time. Then, we replaced MACR with DCR and observed significant (p<0.05) additional gain of 0.85\% and 1.53\% for effective VV and video watch time respectively. Furthermore, we classified the items as long-tail (<5,000 impressions), hot (>180,000 impressions), and others based on their exposure count before the A/B testing. We observed 4.69\% more exposure for long-tail items, less exposure of 1.01\% for hot items, and marginal change for the other items by substituting DCR for MACR. 

\section{Conclusion}
In this paper, we propose a structural causal model-based method, named \textbf{DCR}, to counteract the popularity bias with disentangled counterfactual reasoning in the sequential recommendation and achieve superior recommendation performance. We conduct thorough offline experiments to prove DCR significantly enhances the recommendation relevance over the existing debiasing methods on multiple real-world datasets and across various sequential base models. Moreover, extensive ablation studies are carried out to verify and understand the effectiveness of each newly introduced component. Finally, DCR achieves significant online gains over the SOTA debiasing methods based on the A/B tests.  

\bibliographystyle{ACM-Reference-Format}
\balance
\bibliography{references}

\end{document}